\documentclass{emulateapj}
\usepackage{./apjfonts}

\usepackage{epsfig}

\def\P3hat{{\mathaccent 94 P}_3}

\def\uv{{\it u-v~}}
\def\ie{{\it i.e.}}

\newcommand{\lsim}{\raisebox{-0.3ex}{\mbox{$\stackrel{<}{_\sim} \,$}}}
\newcommand{\gsim}{\raisebox{-0.3ex}{\mbox{$\stackrel{>}{_\sim} \,$}}}

\shorttitle{Intrinsic short time scale variability of W3(OH) maser}

\shortauthors{R. Ramachandran, A. A. Deshpande, W. M. Goss}

\begin{document}

\title{Intrinsic Short time scale variability in W3(OH) Hydroxyl masers}

\author{R. Ramachandran}
\affil{Department of Astronomy and the Radio Astronomy Lab, University of 
California, Berkeley, CA 94720; 
ramach@astron.berkeley.edu}
\author{A. A. Deshpande}
\affil{Arecibo Observatory, NAIC, HC-3 Box 53995, Arecibo, PR 00612\\
and\\
Raman Research Institute, Sadashivanagar, Bangalore 560080, India; desh@rri.res.in}
\author{W. M. Goss}
\affil{National Radio Astronomy Observatory, P.O.Box 0, Socorro, NM 87801; 
mgoss@aoc.nrao.edu}

\begin{abstract}
We have studied the OH masers in the star forming region, W3(OH), with
data obtained from the Very Long Baseline Array (VLBA). The data 
provide an angular resolution of $\sim$5 mas, and a velocity resolution of 
106 m s$^{-1}$. A novel analysis procedure allows us to differentiate 
between broadband temporal intensity fluctuations introduced by 
instrumental gain variations plus interstellar diffractive scintillation,
and intrinsic narrowband variations. Based on this 12.5 hours observation, 
we are sensitive to variations with time scales of minutes to hours. 
We find statistically significant intrinsic variations with time scales of 
$\sim$15--20 minutes or slower, based on the {\it velocity-resolved fluctuation 
spectra}. These variations are seen predominantly towards the line shoulders. 
The peak of the line profile shows little variation, suggesting that they perhaps
exhibit saturated emission. The associated modulation index of the observed 
fluctuation varies from statistically insignificant 
values at the line center to about unity away from the line center. 
Based on light-travel-time considerations, the 20-minute time scale of
intrinsic fluctuations translates to a spatial dimension of $\sim$2--3 AU along 
the sight-lines. On the other hand, the transverse dimension of the sources, 
estimated from their observed angular sizes of about $\sim$3 mas, is about 6 AU. 
We argue that these source sizes are intrinsic, and are not affected by interstellar 
scatter broadening.
The implied peak brightness temperature of 
the 1612/1720 maser sources is about $\sim2\times 10^{13}$ K, and a factor 
of about five higher for the 1665 line. 
\end{abstract}


\section{Introduction}
\label{sec-intro}

Interstellar hydroxyl maser sources are found in the Galaxy and many
external galaxies. In the Galaxy, these sources are associated with star
forming regions (predominantly in the 1665 and 1667 MHz lines), IR and
late type stars (mainly at 1612 MHz) and supernova remnants
(exclusively at 1720 MHz). 

Numerous VLBI observations (e.g., Gwinn et al 1988; Fish et al 2005; 
2006) have indicated
that the apparent angular sizes of OH masers increase with distance in the Galaxy. 
OH maser sources in the inner Galaxy show larger angular
sizes compared to sources in the outer Galaxy. These facts 
strongly suggest that the observed angular size is strongly dependent on
interstellar scatter broadening. Some years ago, Burke et al (1968)
and Gwinn et al (1988) have proposed that the apparent broadening of masers shows a
wavelength dependence $\propto\lambda^2$, based on OH maser sizes at
1665 MHz and $H_2O$ maser sizes at 23 GHz. These data suggest that 
interstellar scattering is  the dominant cause for the observed
angular sizes of these sources.

Desai, Gwinn \& Diamond (1994) used the VLBA to study the details
of the interstellar broadening of the OH masers in the distant 
($\sim$14 kpc) HII  region W49N. Anisotropic broadening was observed with
the orientation of the minor axis preferentially parallel to the
Galactic plane, an additional confirmation of the influence of
interstellar scattering on the scattered sizes. An unambiguous
determination of the intrinsic sizes of OH maser sources would
represent an important input in understanding the OH maser pump and
emission mechanism.

In addition to the intrinsic sizes, an important property of the
masers that provides information about the pumping mechanism is the
variability of the maser intensity as a function of time. Variability on long
time scales (weeks to months) has been discussed by numerous investigators 
(Schwartz, Harvey \& Barrett 1974; Coles, Rumsey \& Welch 1968; 
Zuckerman et al 1972; Rickard, Zuckerman \& Palmer 1975; Gruber \& de 
Jager 1976; Clegg \& Cordes 1991). The suggestion was made that the observed
variability may arise from  changes in the number density of molecules 
exhibiting maser phenomenon, or physical conditions in the maser column. In contrast to
the long timescale variations, short timescale variations
have also been detected (Zuckerman et al 1972; Rickard et al
1975). Salem \& Middleton (1978) provide  a model consisting of a sudden onset
of a pumping mechanism that could  cause rapid quasi-periodic
fluctuations in the observed intensity. The predicted fluctuations on
time scales of roughly a day would have a $\sim$25\% modulation index. 

Evans et al (1972) have investigated 
eight well known OH masers with the goal of describing the
statistics of the radiation. They sampled the output of the 140 feet
NRAO telescope rapidly and find that the radiation is of Gaussian 
nature to within a level of about one per cent.

The major existing study to date of short term variations consists of
Arecibo (beam size of 3 arcmin) and VLA (beam size of 15$\times$5 arcsec)
observations probing variability on 
time scales in the range 16 s to 2 hours (Clegg \&
Cordes 1991). Typical variations were detected at the 5-10 per cent level with 
some large variations at the 100 per cent (or more) level for the sources W75S 
and NGC 6334F. 
As Clegg \& Cordes point out, an identification of any
intrinsic variability in these sources would provide a unique
opportunity to determine the source extent using light-travel-time
arguments. They do find prominent variations in some  of the sources
with time scales of $\sim$20 minutes (e.g S269). However, it is 
very difficult to distinguish between variations that are intrinsic to the 
source, and those that might arise from interstellar diffractive scintillations.
These authors do not decisively choose between an intrinsic or scattering 
origin for the fluctuations. However, they do stress that if the fluctuations 
are due to interstellar scintillation, implausible brightness 
temperatures would be required.

As pointed out by Clegg \& Cordes, several aspects can affect 
their conclusions: 
\begin{itemize}
\item The 3 arcmin beam size of Arecibo,
or even the 15$\times$5 arcsec beam of VLA imply that a number of 
individual maser sources are observed simultaneously. Thus the observations 
can consist of incoherent superposition of multiple sources. 
\item The velocity resolution of the VLA was
only 1.1 km s$^{-1}$; thus in some cases multiple velocity components 
may exist in a single velocity channel since a spectral resolution at the level of
$\lsim$0.1 km s$^{-1}$ is required to resolve complex OH maser
lines. 
\item The total time span of each observation was only $\sim$ 2 hours, thus limiting
the information concerning  intensity variations to time scales of about an
hour.
\end{itemize}

In the current study, we have circumvented these problems by using 
data from the
Very Long Baseline Array (VLBA) of the National Radio Astronomy 
Observatory\footnote{The National Radio Astronomy Observatory (NRAO) 
is a facility of the National Science Foundation operated under a 
cooperative agreement by Associated Universities, Inc.} to produce time
series of OH observations at all four spectral lines of OH in the
source W3(OH). With a total angular extent of the maser region of
$\sim$ 2 arcsec, many maser spots are observed simultaneously. With the $\sim$ 5 mas
spatial resolution and a spectral resolution of $\sim$ 0.1 km s$^{-1}$,
the confusion problem is solved since strong isolated maser sources
can be observed at unique positions and velocities in three of the
four OH lines. With the snap-shot capability of the VLBA, images are
made at intervals of 1 min over a 12.5 hour period. The
modulation indices of the maser lines are derived as well as the power
spectra. After summarizing the observations in \S\ref{sec-observations}, 
our analysis procedure is described in \S\ref{sec-procedure}. 
In \S\ref{sec-variations} we discuss the possible contributors to 
apparent intensity variations, and describe our model incorporating their
spectral characteristics.  
We have adopted 
two methods to explore the intrinsic variabilities in these maser sources;
these are summarized in \S\ref{sec-corranal} and \S\ref{sec-vrspec}.
As we describe in these sections, we find significant intrinsic variations.
Implications of these variations are described in the last section, 
\S\ref{sec-discussion}. Throughout this paper we assume that the
distance of W3(OH) is 2.0 kpc based on the recent VLBA parallax
determinations of Xu et al (2006) and Hachisuka et al
(2004). 

\section{Observations}
\label{sec-observations}
The dataset is the observations analyzed by 
Wright, Gray \& Diamond (2004a,b and 2005; hereafter WGDa,b,c). They 
observed W3(OH) on 1996 August 2 using the VLBA in all the four ground-state lines 
(1612, 1665, 1667, \& 1720 MHz), simultaneously. We have obtained the
calibrated \uv data from Phil Diamond and analyzed the data in order to
search for short term time variations. The data were recorded with
full polarization
information with a bandwidth of 62.5 kHz at all four OH maser
lines. The assumed rest frequencies of these four spectral lines were
1612.231, 1665.402, 1667.359, and 1720.530 MHz, respectively. With 128
spectral channels, the channel separation is
488 Hz, and the resolution of 586 Hz corresponds to a velocity resolution of
0.11 km s$^{-1}$ at 1665 MHz. 

A full description of the adopted calibration procedure can be found in WGDa.
For the purpose of calibration, the sources 3C84 and J1611+343 were
observed during the run. In particular, the amplitude calibration was
carried out using the VLBA parameters (gain curves and system
temperature determinations) known {\it a priori}, and in addition using the 
``template fitting'' method using the AIPS task ACFIT. The overall flux density
is thus determined with an accuracy of a few percent. The phase
calibration was carried out using a phase reference velocity 
channel at 1665 MHz at the velocity of --47.46 km s$^{-1}$ from
W3(OH). This calibration (at the sub-minute time scale) was carried out using
FRING in AIPS followed by a self-calibration in order to remove any
effect of source structure. No amplitude self-calibration was
performed. The phase corrections were then applied to all channels.
In Figure 1 the total \uv coverage of the data is shown for the data at 
1665 MHz. 

\begin{figure}
\begin{center}
\epsfig{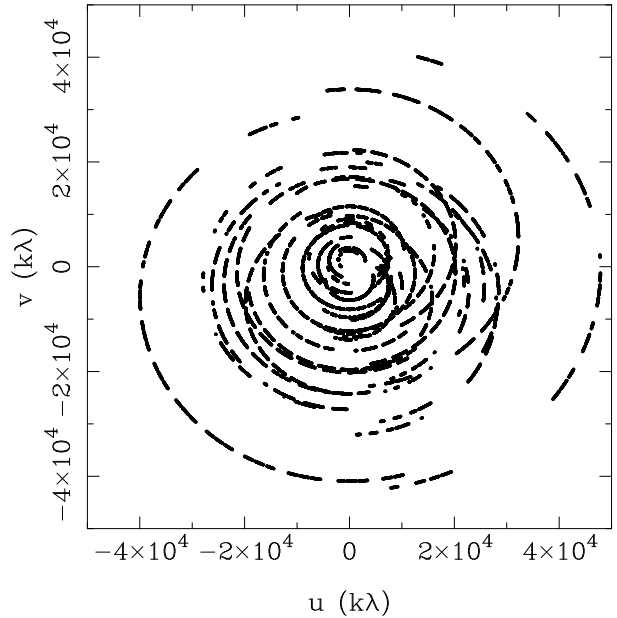}
\caption[]{The total $u-v$ coverage of the data set at 1665 MHz. The 
two axes are in units  of kilo-wavelengths.}
\label{fig:uvfull}
\end{center} 
\end{figure}

\section{Analysis Procedure}
\label{sec-procedure}
The first step in the data analysis is to identify bright sources that 
are spatially well isolated. At the end of this step, three
sources were chosen, whose properties are summarized in Table 1. The spectra from each 
of the three positions  are shown in Figure 4 for the 1612, 1665 and 1720 
MHz lines based on the 12.5 hour observation. At 1667 MHz no strong isolated 
source could be identified that did not exhibit a pronounced velocity 
gradient in adjacent channels. We have not analyzed the 1667 MHz data since 
the gradients may be the result of blended sources. The three positions 
and six lines (RR and LL for each) shown in Table 1 were chosen for 
further analysis.

\begin{table}[h]
\vspace{0.1cm}
\begin{center}
\begin{tabular}{llccl}\hline
{\bf No.} & {\bf line} & {\bf $\alpha$ J2000} & {\bf $\delta$ J2000} &
{\bf Pol} \\
  &   &  {\tt [hh:mm:ss.s]} & {\tt [dd:mm:ss.s]} &  \\ \hline\hline
1 & 1612 & 02:27:03.818 & +61:52:24.439 & LL \\
  & 1612 & 02:27:03.818 & +61:52:24.439 & RR \\
2 & 1665 & 02:27:03.825 & +61:52:24.653 & LL \\
  & 1665 & 02:27:03.825 & +61:52:24.653 & RR \\
3 & 1720 & 02:27:03.829 & +61:52:24.704 & LL \\
  & 1720 & 02:27:03.829 & +61:52:24.704 & RR \\ \hline
\end{tabular}
\end{center}
\label{tab:srclist}
\caption{List of maser spots used for the present analysis. The second column
gives the line tag in MHz, the third and the fourth give the right ascension
and declination (J2000) of the sources, along with the circular polarization tag
indicated in the last column.}
\end{table}

The positions for the six lines were obtained after correction of the
WGDa,b positions. Since the August 1996 data is not phase referenced
and is self calibrated, the absolute coordinates are uncertain at the
0.1 arcsec level. The absolute positions were tied by WGD a,b to
earlier determinations of absolute positions by Gray et al (2001) with
an uncertainty of $\sim$10 mas. We have used the corrections to the
WGDa,b positions published by WGDc. In addition, in December 2004,
Palmer and Goss (unpublished) have carried out phase referenced VLBA
observations with an astrometric precision of $\sim$1 mas. These
authors have determined that the corrections to the WGDa,b coordinates
are $+48\pm 5$ mas in right ascension (\ie ~add 48 mas of arc to the
coordinates published by WGDa,b) and $+184\pm 5$ mas in declination. 
WGDc have suggested that these values are $+54$ mas and $+175$ mas,
respectively. Both determinations are in reasonable agreement. The coordinates
in Table 1 are the WGDa,b values using the corrections determined 
by Palmer and Goss.

\begin{figure}
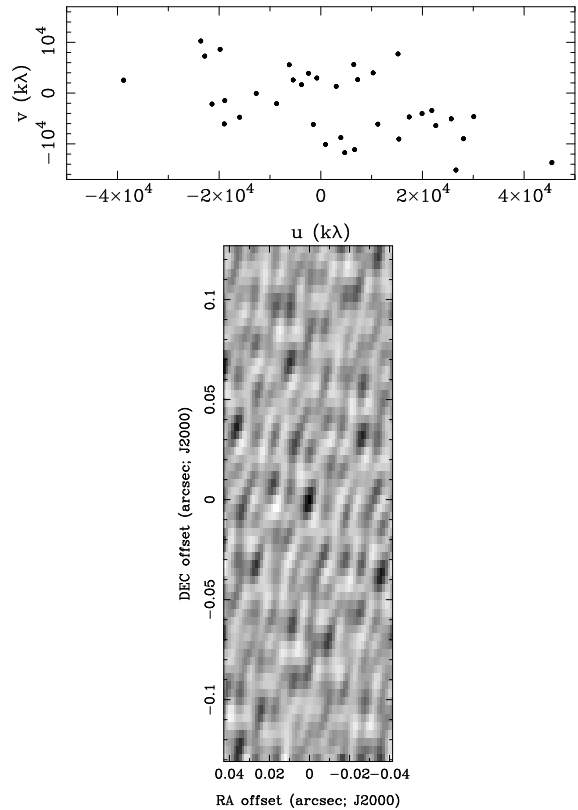

\begin{center}
\epsfig{file=fig2a.ps,height=7.5cm,angle=-90}
\vspace*{0.7cm}
\epsfig{file=fig2b.ps,width=7.5cm,angle=-90}
\caption[]{{\it Top:} Typical \uv distribution with 1-min
integration, corresponding  to the 1665 MHz observation at an hour angle of
4 hours. {\it Bottom:} Typical one minute map produced with the 1665 MHz data.}
\label{fig:uv1min}
\end{center}
\end{figure}


\begin{figure}
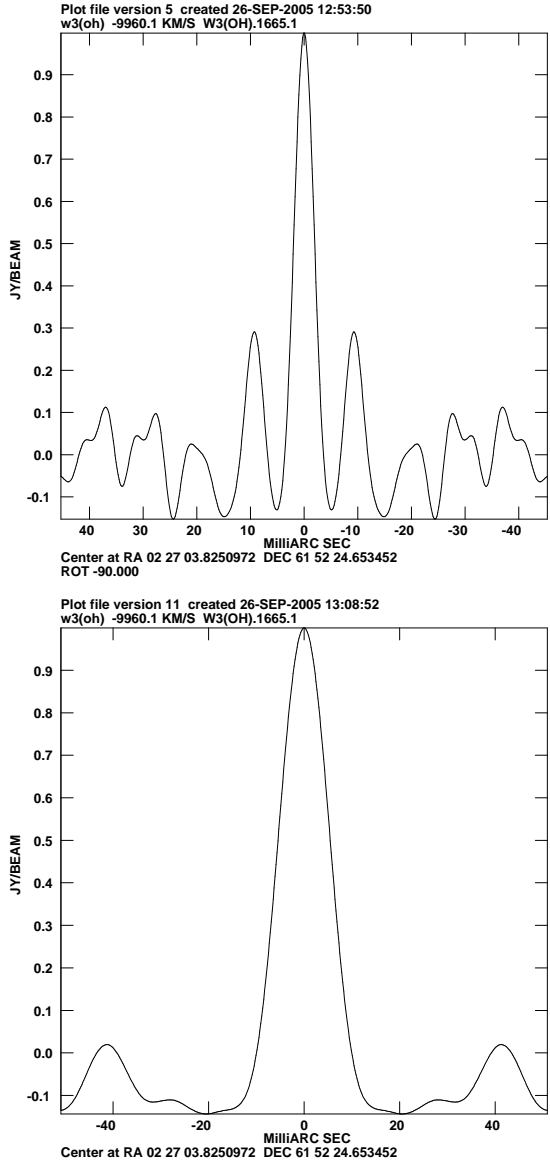

\begin{center}
\epsfig{file=fig3a.ps,width=7.5cm}
\vspace*{0.7cm}
\epsfig{file=fig3b.ps,width=7.5cm}
\caption[]{{\it Top:} Beam pattern as a function of right 
ascension and declination are given in the top and the bottom 
panel, respectively, for a data of one minute duration.}
\label{fig:beam1min}
\end{center}
\end{figure}

\begin{figure}
\begin{center}
\vspace*{0.7cm}
\epsfig{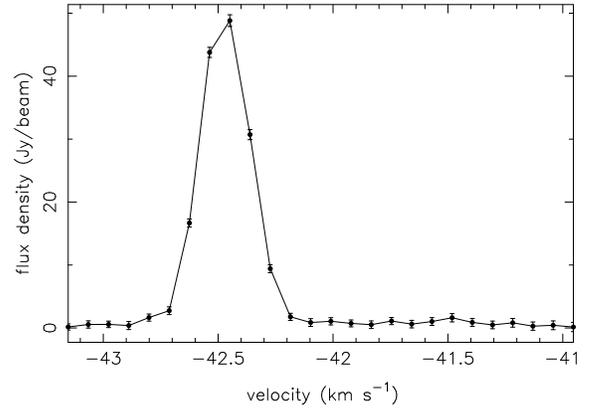}
\caption[]{Spectral line profile of W3(OH) corresponding to the one
  minute integration at 1665 MHz.}
\label{fig:prof1min}
\end{center}
\end{figure}

Numerous trial snap-shot images were made in order to determine the
minimum time interval over which a successful and reliable image can
be obtained. For these isolated strong sources
($\sim$10 Jy beam$^{-1}$), we find 1 minute as the shortest interval at
which reliable images can be constructed with minimal problems of
confusion (see below for a special case for the 1612 MHz line). In
the top panel of Figure \ref{fig:uv1min}, we show the \uv coverage of 
a one minute interval at an hour angle of 4 hours for the 1665 MHz 
observation. To give an idea of the quality of map, we also show
the map produced by a one minute observation in the bottom panel.
Moreover, in the top and the bottom panels of Figure \ref{fig:beam1min}, 
we show the cross 
section of the beam along the right ascension and the declination 
directions, respectively. In order to demonstrate the signal to noise 
ratio of the 
detection of the spectral lines, we also show in Figure \ref{fig:prof1min}
the spectrum of the 1665 MHz line corresponding to the 1-min integration.

\begin{figure*}
\begin{center}
\epsfig{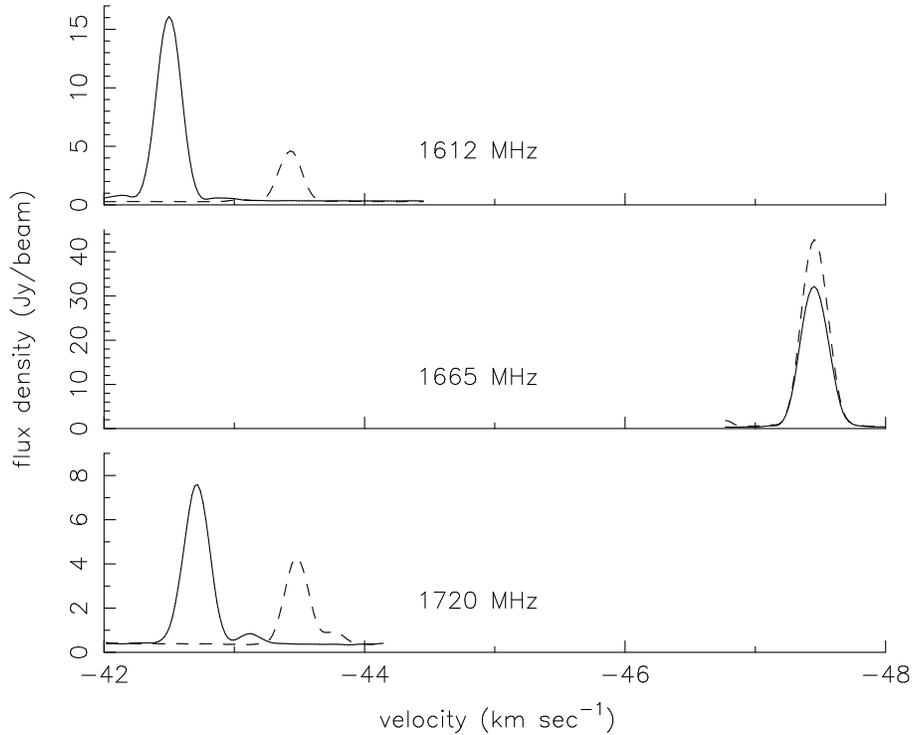}
\caption[]{Integrated line profiles of W3(OH)corresponding to the three sources 
summarized in Table 1. Profiles with solid lines correspond to LL 
polarization, and the ones with `dashed' lines correspond to the 
RR polarization.}
\label{fig:zeeman}
\end{center}
\end{figure*}

Out of the three sources given in Table 1, the sources at 1612 and 1720 
MHz have been identified as Zeeman pairs. The corresponding longitudinal 
magnetic field strengths are given in the table from WGDa. At 1665 MHz, 
we do not observe 
a statistically significant shift between the line profiles in RR and 
LL channels. The integrated (over 12.5 hours) line profiles of these 
Zeeman components are shown in Figure \ref{fig:zeeman}.

The data analysis to identify and compute the line profiles was carried out using 
MIRIAD. Since some of the sources are clearly Zeeman pairs with velocity 
separations of the order of $\sim$1 km s$^{-1}$, (several freq channels), we 
carried out the analysis in both  RR and LL polarization channels independently.
The rationale for this procedure is that the intensity variations
observed in the two polarizations may not necessarily be correlated
even for small (< 0.1 to 0.2 km s$^{-1}$) velocity separations. 
The full 12.5 hour observation was used to 
constrain the source position and shape parameters (major and minor axis and 
PA). Then, we determined the flux density at 1 minute intervals, with the 
source position and the shape parameters held fixed. For this purpose, the 
MIRIAD task UVFIT was used. At the end of this step, we obtained line profiles
every minute in RR and LL channels separately for all sources listed in Table 1. 
These profiles were arranged chronologically, forming a dynamic spectrum
of intensity as a function of time and frequency (or velocity), 
for each source, in RR and LL separately.

For the 1612 MHz line, it was necessary to carry out a minor correction due to 
source confusion. The major 1612 MHz line has flux densities of 15 Jy 
(RR) and 4 Jy (LL). There is another source nearby (displaced by $-12$ mas 
in right ascension and $+1$ mas in declination) in the same velocity range.
The flux density of this source is $\sim$1 Jy, which is about 10\% of the more
intense source. This weaker confusing source 
was  imaged using  the 12.5 hour observation; the mean flux density 
was subtracted from the entire \uv data base before the time series 
was constructed for the two major lines at RR and LL polarization. 
This method is based on the assumption that the time variation of the 
confusing source is minor compared to that of the brighter source
based on the relative weakness of the confusing source. 

\section{Origin of flux density variations, and our model}
\label{sec-variations}
Flux density 
variations with time scales of tens of minutes can be introduced by 
various effects that are independent of intrinsic variability. In order
to explore the possibility that these maser sources exhibit 
intrinsic variability, it is essential to understand 
the nature of additional  {\it extrinsic} sources of variability
on relevant time scales.

First, any uncorrected instrumental gain variations in the receiver system will 
introduce apparent intensity variations. However, these changes will be 
correlated across spectral channels. In other words, we expect 
these changes to have a correlation bandwidth that is far wider than 
the typical spectral line widths of the maser sources of $\sim$0.5 km s$^{-1}$.

There is an effect that can potentially introduce flux density
variability on short time scales. This effect arises  from  interstellar 
diffractive scintillations; two of the important parameters characterizing 
it are, namely, the decorrelation bandwidth ($\Delta\nu$), and the 
diffractive scintillation time scale ($T_{\rm dif}$). The latter ranges 
typically from minutes to hours. If the decorrelation bandwidth value is 
comparable to, or narrower than, the spectral line width of the maser source,  
i.e. less than a few channel widths or less than 0.2 to 0.4 km s${-1}$, then 
we would consider the temporal variations introduced by diffractive 
scintillations to be ``narrowband'', and  would expect significant differential 
variation within the line profile. In order to assess the situation correctly, 
we must estimate the expected value of the decorrelation bandwidth of 
diffractive scintillations along this sight-line.

Since these maser sources are spectral line emitters, measurements
of the decorrelation bandwidth are often uncertain. However, the 
measured angular diameters of these sources provide an indirect
estimate. The apparent angular width of many of the  sources
in W3(OH) are measurable in this data set, and is typically $\sim$3 mas
(Palmer \& Goss, Private communication). 
If we  assume that this width is exclusively dominated by
interstellar scatter broadening, i.e. the intrinsic angular size
is significantly smaller than the apparent size,  
it is possible to calculate the
expected {\it relative time delay} associated with the scattered
rays with respect to the direct path (Gwinn et al. 1993; Deshpande \&
Ramachandran 1998). This effective time delay is identical to
the {\it characteristic temporal broadening scale} ($\tau_{\rm sc}$) in 
the case of pulsar pulse profiles. At a distance of 2.0 kpc, and assuming that the
scattering material is uniformly distributed along the line of sight, the delay is
\begin{equation}
\tau_{\rm sc}
= \frac{D\;\theta_H^2}{16\;c\;\ln 2}\;\sim 4.5\;\; \mu{\rm s,}
\end{equation}
where $D$ is the distance to the object, $\theta_H$ is the angular
width of the source (full-width at half maximum), and $c$ is the speed
of light. Since the decorrelation bandwidth and the effective time delay obey the 
following ``reciprocal" relation,
\begin{equation}
2\pi\tau_{\rm sc}\;\delta\nu\;\approx\;1,
\end{equation}
the value of the decorrelation bandwidth, $\delta\nu$ is then
predicted to be $\sim$35 kHz. This is significantly greater than the typical
line widths (1-2 kHz, or 0.2-0.4 km s$^{-1}$). Of course, this
estimate is a worst-case estimate, 
where we have assumed that the observed width entirely arises from  
interstellar scattering. Even with this worst-case estimate, the 
expected differential variation within the spectral line profile
as a function of time is only about $\sim$1\% or less.

Apart from the above mentioned causes for intensity 
fluctuations, any additional observed variations may be assumed to be 
intrinsic to the maser source. Of course, we have no {\it a priori} 
knowledge about the nature of  the intrinsic variations in OH masers. However, for the current 
analysis, we assume that the correlation bandwidth of any intrinsic variations is
small enough to consider them as  ``narrowband''. That is, such variations in 
one channel would, in general, be uncorrelated with those in the other channels 
separated by our velocity/spectral resolution or more. There is an important reason 
for making this assumption. The possible intensity variations due to instrumental 
gain instability and interstellar diffractive scintillations are expected
to be correlated across a velocity/spectral range much wider than the maser 
line widths, and hence can be treated as ``broadband'' variations, i.e. as 
modulations that are common to all the observed spectral/velocity channels.
Therefore, in our analysis, we will be sensitive to only those intrinsic variations 
that are narrowband in nature, thereby clearly distinguishable from the instrumental
and interstellar effects.

In order to quantify the observed variations, 
we have adopted the following procedure.  We make a distinction between
the different contributors to the observed flux density variance,
\begin{equation}
\sigma^2_{\rm obs}(v) \;=\; \sigma^2_b + \sigma^2_i(v) + \sigma^2_n(v).
\end{equation}
The three terms in the above equation result from a ``broadband'' modulation, 
possibly intrinsic narrowband variability, and measurement uncertainty, 
respectively. For this purpose, we model the observed intensity $S_{obs}(t,v)$,
i.e. a function of both time ($t$) and velocity ($v$) as
\begin{equation}
S_{\rm obs}(t,v) = S_{\rm ave}(v)\;[1 + f_b(t)]\;[1 + f_v(v,t)]\;+\;n(v,t),
\end{equation}
where $f_b$ and $f_v$ are the zero-mean fractional variations that
are of ``broadband'' and ``narrowband'' nature, respectively, and
$n$ is the measurement noise, assumed to have zero mean. 
The function $S_{\rm ave}(v)$ refers to the average line-profile
as a function of velocity. The time averaged cross-correlations between 
$f_b$, $f_v$ and $n$ are expected to be zero, i.e. they are assumed to 
be mutually uncorrelated. The parameter $f_b$, the ``broadband'' modulation,
is independent of velocity (or frequency), and is therefore 
described as a function of time alone. 
Adopting the above formulation,
the observed variance $\sigma^2_{\rm obs}$ in a given velocity channel
can be expressed as
\begin{equation}
\sigma^2_{\rm obs}(v) = S^2_{\rm ave}(v)\;[\sigma^2_c + \sigma^2_v(v)] + \sigma^2_n(v),
\end{equation}
where $\sigma_c$ and $\sigma_v(v)$ are the root-mean-square ({\it RMS}) fluctuations
characterizing the fractional temporal variations $f_b$ and
$f_v(v)$, respectively. $\sigma_n(v)$ is the {\it RMS} uncertainty in
the measurements, which can be estimated in a straight-forward 
manner as $\sqrt{\sum e^2(v,t)/N}$ or $\sqrt{\langle e^2(v,t)\rangle_t}$.
Here, $\langle x\rangle_t$ represents the average of ``$x$'' over time, and
$e(v,t)$ (1-$\sigma$ error), is available for each of the $N$ 
(typically, 300-400) time samples in the
dynamic spectrum $S_{\rm obs}(t,v)$ at a given velocity $v$.
The $e(v,t)$ estimate for relevant channels also includes the contribution
of the maser emission to the system noise.
The desired quantity $\sigma_v(v)$, the modulation index associated
with the ``narrowband'' variation (uncorrelated across velocity 
channels), can be estimated if $\sigma_c$ is known. 

\section{Correlation analysis to estimate ``broadband'' variation}
\label{sec-corranal}
In order to estimate and remove the possible
contribution of any broadband variation (characterized by $\sigma_c$)
from  the observed variance, we examine the
cross-correlations between the intensity fluctuations
in all velocity-channel pairs. Auto-correlations
are excluded since they, as seen from equation 5,
are contaminated by the variance
of measurement noise and narrowband fluctuations.
Since  $f_b$, $f_v$ and $n$ are mutually uncorrelated,
the cross-correlation, expressed as ``cross-variance'', between intensity
fluctuations in any pair of channels (about their respective mean
intensities) is given by

\begin{eqnarray}
\sigma^2_{\rm obs}(v_1,v_2) &=& \langle S_{\rm obs}(t,v_1)\;S_{\rm obs}(t,v_2) \rangle_t \nonumber \\
&=& S_{\rm ave}(v_1)\;S_{\rm ave}(v_2)\;(\langle f_b(t)\;f_b(t)\rangle_t \; + \nonumber\\
&& \; \langle 
f_v(v_1,t)\; f_v(v_2,t)\rangle_t) \; + \; \langle n(v_1,t)\;n(v_2,t)\rangle _t \nonumber\\
&=& S_{\rm ave}(v_1)\;S_{\rm ave}(v_2)\;\langle f_b(t)\;f_b(t)\rangle_t \nonumber \\
&=& S_{\rm ave}(v_1)\;S_{\rm ave}(v_2)\;\sigma^2_c,
\end{eqnarray}
where $v_1\neq v_2$.
In a given velocity channel, the contribution due to any ``broadband''
intensity fluctuations, treated as an amplitude
modulation, is expected to be proportional to the mean intensity
associated with that channel. The cross-correlations
between such fluctuations then will be proportional
to the product of mean intensities of the corresponding
pair of channels. The possible contribution from 
any ``narrowband'' fluctuation, $f_v(v,t)$, will also share this
proportionality, but
the associated cross-correlation is expected to have
a zero-mean value. Similarly, the cross-variance of the measurement
noise in any two different (and independent) velocity channels is also
expected to be zero.  
Based on this assumption, we plot (as in the example shown in Figure 5)
the cross-correlation $\sigma^2_{\rm obs}(v_1,v_2)$ (along the y-axis) against
the associated product of mean intensities 
$S_{\rm ave}(v_1)\;S_{\rm ave}(v_2)$.
From this plot an estimate of $\sigma^2_c$, or the slope of the 
expected linear dependence, can be derived. 
The velocity channels where the signal-to-noise ratio
of the mean intensity
$S_{\rm ave}(v)$ is less than 3 
are excluded from this analysis. Naturally, the
number of useful channels ($M$, typically about 5-6)
based on this criterion is significantly reduced,
but the resultant number of correlation pairs
[$M(M-1)/2$] is large enough.
The observed scatter about the linear dependence arises 
from the terms with $f_v$ and/or $n$, resulting in the uncertainty
in the estimation of its slope. The two straight lines about the
best-fit line, all passing through the origin, indicate the $\pm \sigma$
uncertainty in the slope. With $\sigma^2_c$ estimated in this manner,
the ``broadband'' modulation contribution can now be removed from the 
observed variance. 

In practice, any intrinsic fluctuation which may otherwise be 
uncorrelated between adjacent channels may contribute to some (positive)
correlation due to the finite velocity
resolution of the measured spectra (0.11 km s$^{-1}$ at 1665 MHz). 
Since such contributions will mimic correlations due to ``broadband''
modulation, these contributions can lead to an over-estimation
of $\sigma^2_c$, and as a result,
the contribution from ``narrowband''
fluctuations will be correspondingly underestimated. 
Any intrinsic fluctuation with 
a correlation bandwidth wider than
the channel width will also be underestimated. If these effects are in
fact significant, our estimates of the modulations index (or variance) 
associated with 
intrinsic variability can then be viewed as lower limits. 
As one possible measure against over-estimation of $\sigma^2_c$, 
we limit its value, if needed, such that $\sigma^2_n(v))$ is never negative 
(see Equation 5).

To assess the possible contamination due to these effects, 
as well as the robustness of the correlation procedure,
we have repeated the analysis after normalization of each profile 
in the observed dynamic spectra with respect to
(1) the velocity-averaged intensity, and (2) the intensity in a reference
channel corresponding to
the peak of the average profile. The estimates of
$\sigma_c$ in these two cases are close to zero, suggesting that
the above discussed contaminations are not significant.
Also, the estimates of 
the intrinsic modulation  
are found to be consistent across the three methods, indicating that 
the estimate is not sensitive to whether or how the ``broadband'' 
variation was modified before applying the correlation procedure.

We have also examined the performance of the three methods
when used separately. Although, they provide a 
consistent accounting (and removal) of the ``broadband'' modulation 
contribution, there are significant differences.
Normalization with respect to the peak channel intensity is based on
an  implicit {\it a priori} expectation that 
the intrinsic variability in that reference channel
is {\it absent}, and any observed variability is therefore only 
of ``broadband'' nature.
Invalidity of this expectation can lead to a systematic over-estimation of the
variance in the other channels, depending on the modulation index of
the ``narrowband'' variation (including system noise) in the reference channel.
However, for the channels adjacent to the reference channel sharing any
common ``narrowband'' variability, the resultant variance would be 
underestimated.
On the other hand, normalization based on the velocity-average
intensity makes no assumption about absence of intrinsic variability 
in any particular channel,
but does implicitly assume that the line-integral is intrinsically constant.
Such an assumption has no physical justification.
Hence, any ``narrowband'' variability, apart from its magnitude 
being underestimated, 
contaminates other channels with anti-correlated variations and a 
consequent increase
in the variance. In both of these approaches, the {\it corrected} 
dynamic spectra are available, and can be examined  
using fluctuation spectral analysis for estimating the 
temporal scales  of the remaining variability.
In contrast, the correlation-based estimation and removal of the ``broadband''
contribution to the variance only produces time-averaged quantities,
and hence any further temporal/spectral analysis can not be attempted.
However, the correlation-based approach is the most unbiased in 
comparison with the two methods based on normalization. Therefore, 
for a determination of the variance associated with
the ``broad- and narrowband'' variability, we have used the 
correlation-based procedure described above, and have only employed 
the normalization by intensity in the reference peak channel
for the purposes of temporal/spectral analysis (e.g. \S\ref{sec-vrspec}).

\begin{figure}
\begin{center}
\epsfig{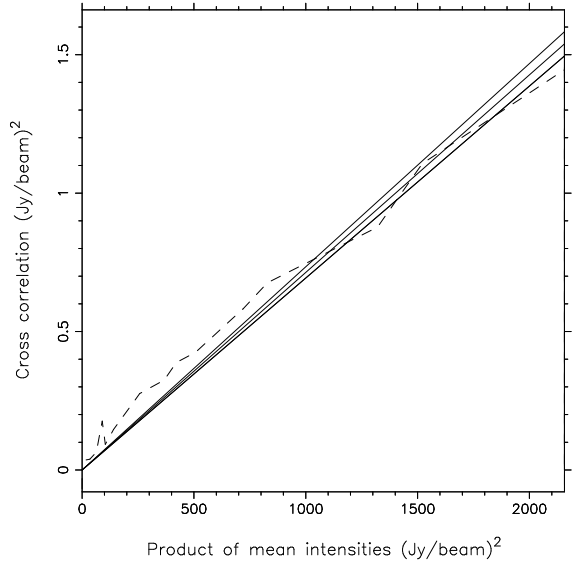}
\caption[]{
An example, based on W3OH 1665-MHz line (RR) data, to illustrate our procedure
to estimate the possible contribution from ``broadband'' modulation.
Here, we examine the cross-correlation between variations observed
in every pair of channels versus the product of mean intensities 
corresponding to the respective channel pair. The observed trend
is shown by the dashed line. Only those channels with
their mean intensities having S/N greater than 3 are included.
The slope of the expected linear dependence provides an estimate of  
$\sigma^2_c$ (see Equation 6), where $\sigma_c$ is the modulation index
associated with the ``broadband'' modulation that is shared by all
channels.
The observed scatter about the linear dependence arises 
in general from any ``narrowband'' variations including the measurement noise.
The middle of the three solid lines, always passing through the origin (0,0),
indicates the best fit to the data, while the two other lines 
correspond to the $\pm 1-\sigma$ uncertainty in the slope.}
\label{fig:xcorr}
\end{center}
\end{figure}

\begin{figure*}
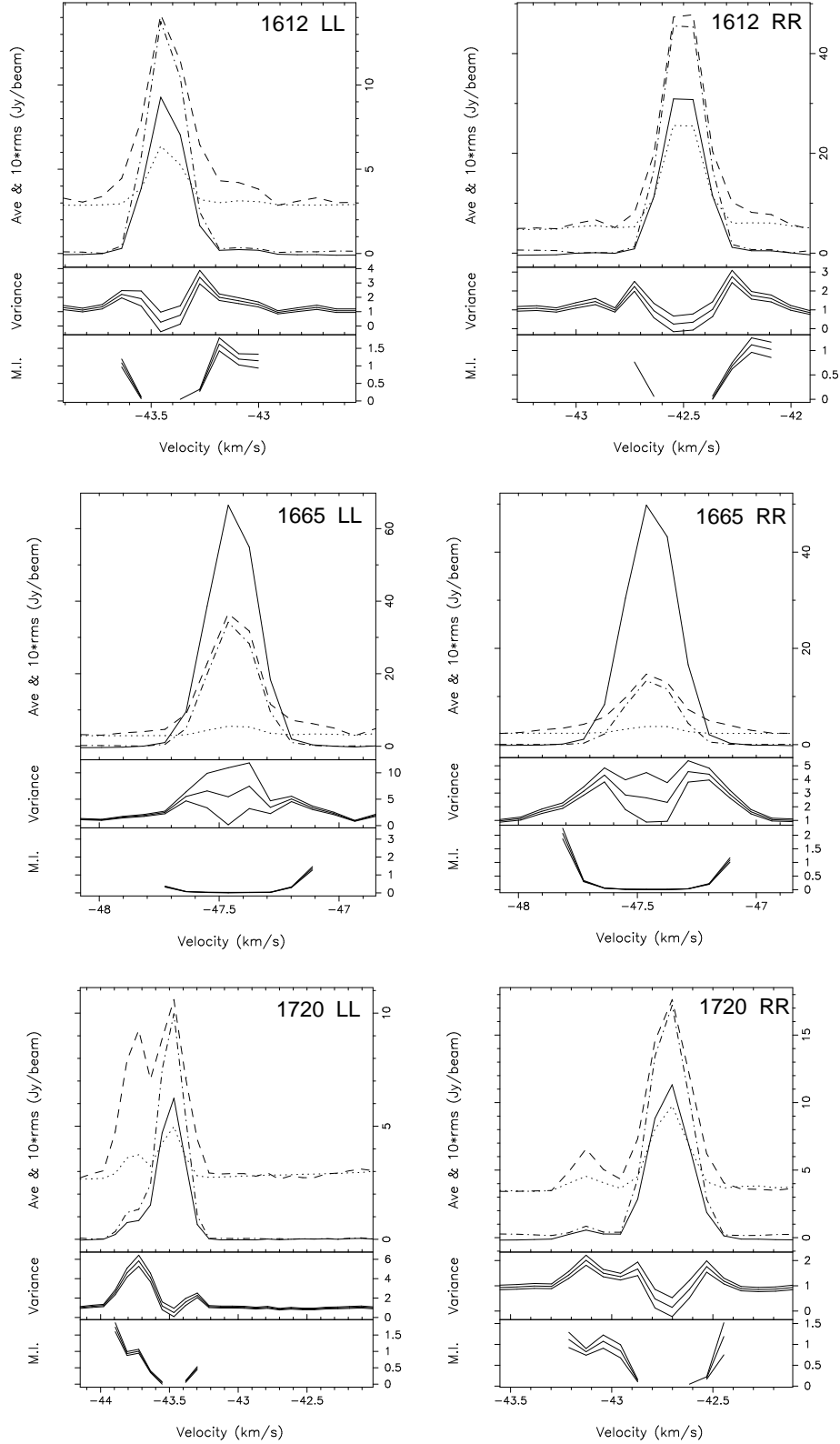

\begin{center}
\epsfig{file=fig7a.ps,height=5.5cm,angle=-90}
\hspace*{1cm}
\epsfig{file=fig7b.ps,height=5.5cm,angle=-90} \\
\vspace*{0.5cm}
\epsfig{file=fig7c.ps,height=5.5cm,angle=-90}
\hspace*{0.5cm}
\epsfig{file=fig7d.ps,height=5.5cm,angle=-90} \\
\vspace*{0.5cm}
\epsfig{file=fig7e.ps,height=5.5cm,angle=-90}
\hspace*{0.5cm}
\epsfig{file=fig7f.ps,height=5.5cm,angle=-90}
\caption[]{Results of the variance analysis of the dynamic spectral data
on the sources listed in Table 1.
The plots corresponding to the LL \& RR 
polarization channels are shown on the left and the right columns, respectively,
The top, middle and the bottom pairs show the results for the 
1612, 1665 and 1720 MHz transitions, respectively.
In each of the six plots, the top panel shows the 
average intensity profile (solid line), along with the profiles of
standard deviations associated with the observed variance (dashed line),
measurement noise (dotted line) and the ``broadband'' modulation
(dot-dashed line), respectively. The latter is a scaled version
of the average intensity profile, where the scale factor $\sigma_c$
is estimated using the cross-correlation procedure in \S\ref{sec-corranal}.
For clarity, all three profiles for the standard deviations 
are amplified by a factor of ten in this display.
The middle panels display profiles of the ratio (R) of the
observed ``narrowband'' variance to its expected value. The observed
``narrowband'' variance is simply the observed variance minus that associated with
the ``broadband'' modulation, while the {\it expected} ``narrowband'' variance
is that from the measurement noise ($\sigma^2_e$). The three curves correspond to
the mean value of the ratio R, and with $\pm 1-\sigma$ deviations from this mean.
The nominally expected values of R is unity within the indicated uncertainties, 
and any significant excess is interpreted as due to 
intrinsic ``narrowband'' variability. The modulation index (MI) associated with 
the latter, along with its $\pm 1-\sigma$ bounds are given in the bottom panel.
The modulation index (MI) is computed as the ratio of the 
{\it RMS} fluctuations attributable to the intrinsic variability to
the mean intensity $S_{ave}$, when both of them are significantly greater than zero.
All of the estimates presented above are based on the data spanning 12.5 hours.}
\label{fig:variance}
\end{center}
\end{figure*}

The results from our variance analysis on all of the 6 data sets 
(consisting of dynamic spectra for RR \& LL polarization for the sources in Table 1) 
are summarized in Figure \ref{fig:variance}, showing in each case the profiles of average
line intensity ($S_{ave}$), as well as the standard deviations associated with the observed
variability ($\sigma_{obs}(v)$), possible ``broadband'' modulation 
($\sigma_b (v)$ = $\sigma_c S_{ave}(v)$) 
and the measurement uncertainties ($\sigma_n$).
As already described and illustrated in Figure \ref{fig:xcorr}, 
the ``broadband'' modulation index $\sigma_c$ is estimated
based on the cross-correlation analysis.   
The quantity $\sigma^2_{obs} - \sigma^2_b$, where the ``broadband'' 
modulation contribution is removed, represents the
observed ``narrowband'' variance, which includes the nominally expected contribution
$\sigma^2_n$ from the measurement noise. Also, we examine the ratio
of the observed to the expected variances, 
$R = (\sigma^2_{obs} - \sigma^2_b)/\sigma^2_n$, 
for any significant deviations from the expected value of unity.
In other words, any statistically significant residual variance, 
i.e. $\sigma^2_i = \sigma^2_{obs} - \sigma^2_b - \sigma^2_n$,
must then be ``narrowband'' in nature, and thus may be assumed to be 
intrinsic to the source. As another measure, we also compute the modulation index 
associated with the possible intrinsic ``narrowband'' variability, 
$\sigma_v = \sigma_i/S_{ave}$, where the signal-to-noise
ratio of $S_{ave}$ estimate is three or more.
The profiles of the ratio $R$ and $\sigma_v$ 
are also displayed in Figure \ref{fig:variance}.

A number of salient facts are evident from Figure \ref{fig:variance}.
Firstly, statistically significant variations are indeed detected, even after
accounting for the ``broadband'' modulation and measurement uncertainties.
These variations, reflected by the excess variance at certain velocities,
are ``narrowband'' in nature, and hence intrinsic to the source. Moreover,
these variations are apparent at velocities away from the line peak,
rather than at the peak. Within the 
statistical errors, the peaks do not appear to exhibit significant variations. 
The variations observed  in the shoulders are detectable with signal-to-noise
ratio of three or more.

\section{Velocity-Resolved fluctuation spectra, and time-scales of intrinsic variability}
\label{sec-vrspec}

The basic data used here also are the dynamic spectra spanning 12.5 hour
and obtained as chronologically arranged line profiles from one minute integrations.
The aim of the fluctuation spectral analysis described below is twofold.
The first is to examine the nature of the fluctuation spectra.
Then, given the fluctuation spectra, to estimate the timescale of flux density 
variations arising from ``narrowband'' intrinsic variations. 
Thus it is essential to eliminate  contribution from any
``broadband'' variations. To ensure this, each of the spectral profiles
in the dynamic spectra is normalized with respective intensities in the
reference channel defined by the peak in the average line profile.
The justification for this normalization comes from the variance analysis 
(\S\ref{sec-corranal}), where we concluded that the peak of the line profile does 
not exhibit any statistically significant ``narrowband'' variation. 
We can also rule out any overestimation of variability in other channels
due to the normalization.

\begin{figure*}
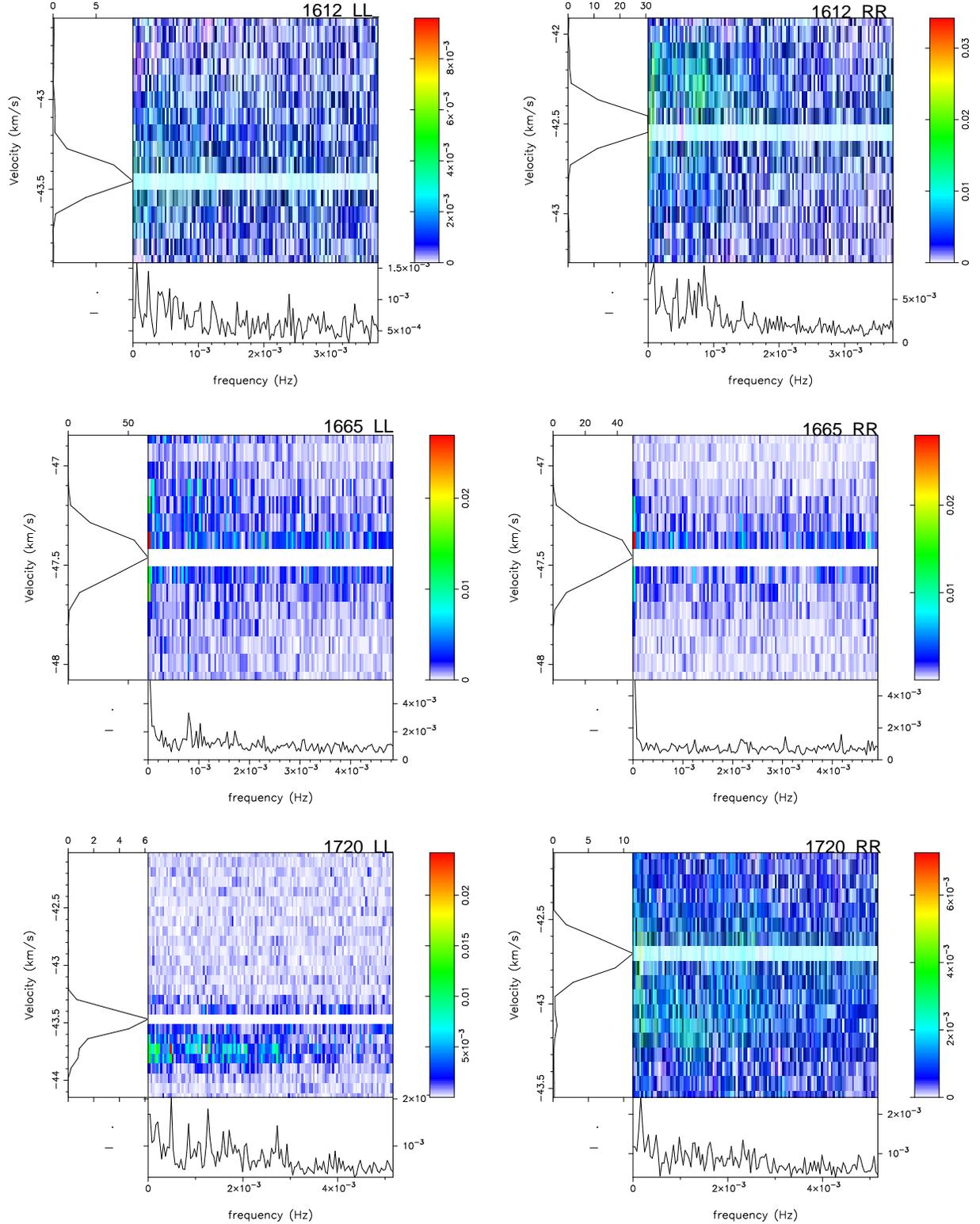

\begin{center}
\epsfig{file=fig8a.ps,height=7.5cm,angle=-90}
\hspace*{1cm}
\epsfig{file=fig8b.ps,height=7.5cm,angle=-90} \\
\vspace*{0.5cm}
\epsfig{file=fig8c.ps,height=7.5cm,angle=-90}
\hspace*{0.5cm}
\epsfig{file=fig8d.ps,height=7.5cm,angle=-90} \\
\vspace*{0.5cm}
\epsfig{file=fig8e.ps,height=7.5cm,angle=-90}
\hspace*{0.5cm}
\epsfig{file=fig8f.ps,height=7.5cm,angle=-90}
\caption[]{Results of the Velocity-resolved fluctuation spectral analysis.
The pair of plots from top to bottom correspond to the 1612, 1665 and 1720 MHz lines,
respectively. In each of these pairs, the left and the right plots are for
the LL, and RR polarization channels, respectively.
Within each plot, the left side panels shows the average line profile
(flux density in units of Jansky/beam) as a function of
velocity obtained from the full observation. 
The central panel shows the velocity resolved fluctuation spectra, where each
horizontal row is the power spectrum of the time series
from the corresponding velocity channel. Each time series is normalized
with respect to the time series for the reference peak channel.
Naturally, this results in a null spectrum for the reference channels.
As the color bar shows, red corresponds to the maximum power, and white
corresponds to zero power. The bottom panel gives the velocity-averaged
fluctuation spectrum. In most cases, significant fluctuation power is
apparent at lower frequencies in comparison with the noise floor at the
higher frequency end of the spectra. As discussed in the main text, the
observed ``narrowband'' variability on time scales of 15-20 minutes 
or longer is to be understood as intrinsic to the maser sources.}
\label{fig:vrspec}
\end{center}
\end{figure*}

These ``corrected'' dynamic spectra for each of the sources form the
input data for the {\it velocity-resolved fluctuation spectral analysis}.
This analysis is very similar to the {\it longitude-resolved fluctuation spectral
analysis}, a well known tool in pulsar emission studies 
(Backer 1973, Deshpande \& Rankin 1999). 
Here, we compute fluctuation power spectrum for each of the velocity
channels separately, where the {\it single} 
temporal sequence of intensities in a given channel is
Fourier transformed and the power at each of the fluctuation frequencies computed.
The results corresponding to all the velocity channels for a given line source
are displayed together as a two-dimensional display of fluctuation power as a function of
velocity and fluctuation frequency, along with a velocity-averaged power spectrum
(see Figure \ref{fig:vrspec}). 

Any statistically significant power observed in these fluctuation spectra can be 
interpreted as due to variability that is necessarily ``narrowband'', and hence, 
intrinsic to the source. The following  aspects are clearly evident from the 
spectra in Figure \ref{fig:vrspec}. The fluctuation power is generally higher in 
the line channels compared to those well away from the line emission, as would 
be expected from the measurement noise that will be proportional to the system 
temperature including the line intensity. Moreover, another thing that is apparent 
is that the fluctuation power toward lower frequencies is more than that at higher 
fluctuation frequencies.  We examine also the spectrum computed by averaging the 
fluctuation spectra across the velocity channels. 
With the equivalent line width of only a couple of channels, given that the contribution 
of the {\it line} channels overwhelms in this average, the resultant 
spectrum has benefited only correspondingly from the averaging across velocity. 
Hence the ratio of the average power to the uncertainties (i.e. the signal-to-noise 
ratio) at any fluctuation frequency is rather small ($\sim$1.4), except when the
fluctuation power in the line channels is small (e.g. as seen
at the high fluctuation frequencies in some of the average spectra). Hence we do not
consider the apparent fine spectral structure as significant,
but rather examine and estimate the smooth trends across the spectrum,
since they will have a much improved significance in accordance with the smoothing
scale.  The fluctuation power level at higher frequencies is consistently low, and
corresponds largely to the measurement-noise contribution that is generally
expected to be ``white'' (or uniform) in its spectral character.
Thus the overall increase in the fluctuation power toward
the lower frequency portion is significant in most cases (except for 
the 1665 line), where the power is typically about two to three times that
at the higher frequency end of the average spectrum.
The latter defines the reference noise floor, across which
the {\it RMS} variation may be estimated and used to assess the
significance of the contrast of fluctuation power levels between the two
frequency ranges. For example, the increase in the average power
toward lower frequencies for the 1612-MHz RR line is about seven times the
{\it RMS} deviation in the power at the higher frequencies. This factor is
somewhat lower for other lines/polarization channels, and is close to
zero for the 1665 RR line. 

We observe significant relative fluctuations up to a fluctuation 
frequency of $\sim 10^{-3}$ Hz, which corresponds to time scales of 
$\gsim$15--20 min. Given that the normalization procedure has removed any 
extrinsic variability that is expected to be ``broadband'', we associate this 
observed time scale with the intrinsic variability of the OH maser lines.

\section{Discussion and Summary}
\label{sec-discussion}

In this work, we have conclusively demonstrated the 
presence of ``narrowband'' intrinsic variations in these W3(OH) maser sources.
These variations seem to have a typical time scale of about 15-20 minutes
or longer, indicating that the faster time scale
may correspond to the light travel time of the maser pumping column. 

A very important aspect, which is worth stating again, is that the 
observed intrinsic variations that are clearly separable from other
contaminants are necessarily ``narrowband'' in nature. 
All the other variations such as the instrumental gain variations
and the interstellar diffractive scintillation, although their time
scales may be of the order of a few tens of minutes, are
distinguishable from the intrinsic
variabilities, 
mainly because of their broad correlation bandwidth. The expected differential 
fluctuation within the line profile due to interstellar scintillation 
is only of the order of $\sim$0.5\%, much smaller than the observed intrinsic 
variability, whose magnitude was as much as 100\% (modulation index), in the ``tail"
portions of the spectral lines.

There is one source of systematic narrowband error that could have potentially 
influenced our conclusions. In our VLBI measurements, if the Earth's 
motion were not correctly compensated for, certain systematic but spurious 
temporal variations in the line profiles would be induced due to differential 
Doppler shift. However, we would then expect the variations seen on the two sides 
of the line profile peak to be anti-correlated. We have examined carefully the 
relevant cross-correlations, and do not find any statistically significant 
signature such an anti-correlation, clearly indicating that our data are free of 
such artefact.

An important aspect of these maser sources is that their typical
measured angular diameter is $\sim$3 mas. With the distance to the
star forming region of 2 kpc, this corresponds to a transverse
distance scale of $\sim$ 6 AU. As we have seen from Figure
\ref{fig:vrspec}, the fluctuation frequencies seen in the spectra are
$\lsim 10^{-3}$ Hz (fluctuation time scale, $T_f\sim 1000$ s or longer). 
If indeed this time scale reflects the dimension of the
source based on light travel time arguments, then the implied longitudinal
spatial scale of the maser column would be 2--3 AU. 
It is important to note that the apparent transverse spatial
dimension of the source as measured from the angular size
is comparable to its longitudinal dimension
implied by the fluctuation timescale, even though there seems to
be no {\it a priori} basis for the comparison, let alone the agreement.
However, if indeed these orthogonal dimensions of the source 
are expected to comparable, it would imply that the
VLBA observations may have actually resolved the intrinsic source size of the OH
masers in W3(OH). Then the possible contribution of
angular broadening caused by interstellar scattering to
the apparent size of the source is minimal. This
also implies that the expected decorrelation bandwidth of
interstellar scintillation may be much wider
than 100 kHz, further reinforcing the validity of our assumptions.
However, if the scatterer is closer to the source instead of the midway location
which is implicit in equation 1, the decorrelation bandwidth would be correspondingly
narrower. We assess whether that may be the case,
based on the two important estimates we have at hand, namely
   the measured upper-limit for scatter broadening $\theta_H$ and the
   observed time scales (say, $T_{\rm dif}$) of the narrow-band variability, if due
   to ISM, since that is our concern.
   How far are we from the naive assumption of uniform scatterer (or a
   strong scatterer mid-way) will be reflected in both of these parameters,
   which depend differently on the relative location of the scatterer
   (see for example, eq. 3 and 8 of Deshpande \& Ramachandran 1998).
   If $D_s$ \& $D$
   are the distances of the scatterer and of the observer, respectively, as measured
   from the source, it is easy estimate an upper limit to   
   ($D/D_s$ -1) using relevant expressions, and 
   assuming that the observed time-scales are 1000 s
   or longer, \& a wavelength of 18 cm. We find that 
   ($D/D_s$ -1) should be less than or equal to $20/(\theta_H V_{ism})$, 
   where the scatter broadening $\theta_H$ is in mas, and 
   the ISM velocity $V_{ism}$ is in km s$^{-1}$. 
   A value of 3 mas for $\theta_H$ implies $(D/D_s -1) \le 
   6.6/V_{ism}$. Noting that $V_{ism}$ will have the same lever-arm
   factor as source velocity does, and considering typical values of
   $V_{ism}$, we conclude that $(D/D_s -1)$ is close to
   unity, if not smaller. Hence, the decorrelation bandwidth
   that we estimate naively is most likely an under-estimate, 
   given the above and that $\theta_H$ may already be an over-estimate 
   of the scatter broadening.

As already mentioned in the introduction, variability of astrophysical masers 
on longer time scales --- weeks to
months --- observed before by several investigators, may be due to changes in the
number density of relevant molecules or physical conditions in the maser
column. On the other hand, very short time
scale rapid variations have also been seen.
Salem \& Middleton (1978) suggest a model in which
a sudden onset of a pumping mechanism can cause rapid
quasi-periodic fluctuations in the observed flux density, and predict
fluctuations with time scales of a day or so, with $\sim$25\%
modulation index. These fluctuations may either correspond to
propagation of a radiative or a collisional disturbance. In the former
case the disturbance travels with the speed of light, and in the
latter with a typical speed of $\sim$10 km s$^{-1}$ or so. For radiative
propagation of disturbance, with the typical flux density of the observed
lines, the brightness temperature comes to $\sim 10^{10}$K. For a
collisional disturbance, the source dimension is only $T_f\times 10$
km s$^{-1}$ $\approx 10^7$ meters (angular diameter of $\sim$0.1
$\mu$arcsec). This corresponds to a brightness temperature of $\sim
10^{20}$K (see also Clegg \& Cordes 1991). In the latter case, since
the implied intrinsic angular diameter is only $\sim$1$\mu$arcsec, the
observed angular diameter is predominantly due to interstellar
scattering. For an object at 2.2 kpc (distance of W3(OH) complex), a
scatter broadening of 2--3 arcsec seems to be very large at the 18-cm
wavelength. VLBI observations of several nearby (distance less than
2--3 kpc) pulsars at 327 MHz have shown a broadening of 10--20 mas,
with an exception of the Vela pulsar, PSR B0833--45 (Gwinn, Bartel \&
Cordes 1993). In the case of Vela, the excess angular broadening is
due to the presence of strong scattering screen (Deshpande \&
Ramachandran 1998). Assuming a Kolmogorov density irregularity
spectrum, with the wavelength dependence of $\lambda^{2.2}$, the
expected angular broadening is only $<0.5$ mas at 1665 MHz.
Based on these considerations, we argue against any
significant over-estimation of the source size of W3(OH) 
due to scatter broadening.

It has been suggested by Elitzur (1991) that short time scale
intensity variations can be produced by variations in the maser level
population at small length scales. These fluctuations produced in the
unsaturated medium of the maser core give rise to the spectrum of flux
density variability observed. In our case, with the length scale of 2-3
AU (derived based on the light travel time arguments) represents an
average ``seed'' length for such a fluctuation. This suggestion may
well be the reason for the short time scale fluctuations that we have
observed.

As mentioned earlier, the longitudinal dimension 2-3 AU estimated from
the observed time scales of intrinsic variability compares well
with the transverse spatial dimension of 6 AU estimated assuming 
that the apparent angular size to be approximately the intrinsic size
of the source. However, for estimating brightness temperature the source
we will use only its transverse size.
A typical observed flux density of some 10 Jy at the line peaks
(e.g. for 1612/1720 lines), implies a peak brightness temperature
of $\sim2\times 10^{13}$ K. A five times higher value,
i.e. $\sim 10^{14}$ K, is implied by the correspondingly 
brighter peaks of the 1665 lines.

The intrinsic intensity variations that we observe are particularly
confined to the ``shoulders'' of the lines and away from the line
peak (see Figure \ref{fig:variance}). The peak of the line profiles does
not exhibit any statistically significant variation, in all of the six data sets. 
This behavior is not at all surprising if the peak of these line emissions corresponds to
``saturated'' maser action, and the rest of the line profile
exhibits unsaturated emission. In any case, the absence of any intrinsic 
variability at the line peak enables use of the corresponding channel
intensities to calibrate out any ``broadband'' variations from the dynamic spectrum.

Another interesting aspect apparent from the results in Figure \ref{fig:variance} 
is that the statistically
significant narrowband variations that we observe at the raising and falling 
edges of the line profiles are not identical between the Zeeman pairs 
(RR and LL components). For instance, the difference between the modulation 
indices of the two Zeeman components of 1720 MHz lines is clearly seen to be several 
times the root mean square noise level. The physical reason for this behavior needs to be 
explored. Given the birefringent nature of the medium close to or at 
the origin of the radiation, as well as interstellar propagation medium, effects 
of refraction on the apparent visibility of the two Zeeman components
remain to be explored.

To summarize, we list our main conclusions as follows.
\begin{itemize}
\item The combination of our cross-correlation procedure and the 
variance analysis provides
an effective tool for estimation and elimination of the ``broadband'' 
contribution from instrumental effects and the interstellar 
diffractive scintillation, and thus for identifying
variations that are intrinsic and ``narrowband'' in nature.
\item Significant intrinsic ``narrowband'' variability is observed
over most regions of the line profiles except at and about the line peak,
suggesting ``saturated'' maser emission at the peak.
\item The velocity-resolved fluctuation spectra reveal that the time scales 
of the significant intrinsic variability are 15-20 minutes or longer.
\item Based on light travel time
argument, the intrinsic variability time scale implies a longitudinal
spatial scale of the maser column to be about 2-3 AU. 
\item The apparent angular sizes of the sources are unlikely to
be significantly affected by interstellar scattering. The implied
transverse size of the source is comparable to its longitudinal dimension. 
\item The peak brightness temperatures for the maser sources range
between  $\sim2\times 10^{13}$ and $\sim 10^{14}$ K.
\end{itemize}

\acknowledgments We thank Phil Diamond for providing us with their
data set for this analysis. We also thank 
Jim Cordes, Vincent Fish and Pat Palmer 
for very valuable discussions and their comments on the manuscript.

\end{document}